\documentclass[preprint,superscriptaddress,aps,floatfix]{revtex4-1}
\usepackage{amsmath}
\usepackage{amssymb, amsfonts}

\usepackage{breqn}
\usepackage{mathtools}
\usepackage{amsthm}
\usepackage{bm}
\usepackage{color}
\usepackage{soul}
\usepackage{breqn}
\usepackage{tikz}
\pdfoutput=1
\usepackage[unicode=true,pdfusetitle,
 bookmarks=true,bookmarksnumbered=false,bookmarksopen=false,
 breaklinks=true,pdfborder={0 0 0},backref=false,colorlinks=true,citecolor=blue]{hyperref}

\usepackage{graphicx}   
\usepackage[caption=false]{subfig}       
\usepackage{etoolbox} 
\usepackage{svg} 

\usepackage{etoolbox}
\usepackage{svg}
\usepackage[nodisplayskipstretch]{setspace}
\usepackage{float}
\usepackage{epstopdf}
\usepackage{cancel}

\makeatletter
\preto\maketitle{%
  \begingroup\lccode`~=`,
  \lowercase{\endgroup
  \let\saved@breqn@active@comma~
  \let~}\active@comma 
}
\appto\maketitle{%
  \begingroup\lccode`~=`,
  \lowercase{\endgroup
  \let~}\saved@breqn@active@comma 
}

\makeatother

\def\be{\begin{eqnarray}}
\def\ee{\end{eqnarray}}
\def\tensorS{\overset{\text{\scriptsize$\leftrightarrow$}}{{\bf S}}}
\def\r{{\bf r}}

\def\E{{\bf E}}
\def\H{{\bf H}}

\def\p{{\bf p}}
\def\m{{\bf m}}

\def\im{{\rm i}}

\def\G{{{\bf G}}}

\def\xim{\boldsymbol{\hat{\xi}}}

\definecolor{JOT-color}{named}{blue}

\begin{document}

\title{Sectoral Multipole Focused Beams}

\author{Jorge Olmos-Trigo}
\affiliation{Donostia International Physics Center (DIPC),  20018 Donostia-San Sebasti\'{a}n, Spain}

\author{Rafael Delgado-Buscalioni}
\affiliation{Condensed Matter Physics Center (IFIMAC), Universidad Aut\'onoma de Madrid, 28049 Madrid, Spain}
\author{Marc Mel\'endez}
\affiliation{Condensed Matter Physics Center (IFIMAC), Universidad Aut\'onoma de Madrid, 28049 Madrid, Spain}
\author{ J.J. S\'aenz}
\email{juanjo.saenz@dipc.org}
\affiliation{Donostia International Physics Center (DIPC),  20018 Donostia-San Sebasti\'{a}n, Spain}
\affiliation{IKERBASQUE, Basque Foundation for Science, 48013 Bilbao,  Spain}



\begin{abstract}
We discuss the properties of pure multipole beams with well-defined handedness or helicity,   with the beam field a simultaneous eigenvector of the  squared total angular momentum and its projection along the propagation axis.  Under the  condition of hemispherical illumination, we show that  the only  possible propagating multipole beams are ``sectoral'' multipoles.  The sectoral dipole beam is shown to be equivalent to the non-singular  time-reversed field of an electric and a magnetic point dipole Huygens' source located at the beam focus. Higher order multipolar beams are vortex beams vanishing on the propagation axis. The simple analytical expressions of the electric field of sectoral multipole beams, exact solutions of Maxwell's equations, and the peculiar behaviour of the Poynting vector and spin and orbital angular momenta in the focal volume could help to understand and model light-matter interactions under strongly focused beams.  
\end{abstract}


\maketitle

\tableofcontents
\section{Introduction} 

``Gaussian beams'' are  the most typical beams employed in optical manipulation \cite{jones2015optical} because most lasers emit light beams whose transverse electric fields and intensity distributions are well approximated by Gaussian functions.
 However, after focusing, when the half width of the beam waist is comparable to the wavelength, $\lambda$, the exact solution of Maxwell's equations predicts significant deviations from the Gaussian shape, only valid in the paraxial scalar approximation (Gaussian approximation can be give accurate results only when the half width of the beam waist is greater than $\sim 10 \lambda$) \cite{varga1998gaussian}.  
 
  A rigorous description of focal and scattered fields, usually based on the multipolar expansion of the field in terms of electric and magnetic multipoles,  is needed for 
 an accurate and quantitative discussion of field intensities, polarization and optical forces near the focal region of a focused beam \cite{sheppard1978electromagnetic,gouesbet1988light,nieminen2003multipole,lock2004calculation,iglesias2011scattering}. A correct multipolar description of the light field is also relevant to analyze 
optical angular momentum (AM)  phenomena \cite{barnett2016optical,barnett2017optical}, including  the interplay between spin (SAM) and orbital (OAM) angular momenta  in the focal volume  of tightly focused beams \cite{bliokh2011spin,bliokh2015spin} as well as to understand  the spectral response  and  optical  torques  on small objects:
Illumination with a tightly focused beam can only excite particle  multipolar modes which are already present in the incident beam \cite{mojarad2008plasmon,zambrana2012excitation}. As a recent example, the strong size selectivity  in optical printing of Silicon particles has been  associated to the dominant contribution of dipolar modes of the tightly focused laser beam \cite{Zaza2019Size}.

 In many optical  applications,  where one wishes to maximize the electric energy density and/or minimize the cross-polarization, a converging electric-dipole wave is the natural choice as the incident wave \cite{bassett1986limit,sheppard1994optimal,dhayalan1997focusing,zumofen2008perfect,gonoskov2012dipole}. 
 since higher order multipoles vanish at the focal point.  In the so-called $4\pi$ illumination (when the incident light can cover the full solid angle), the highest energy density for a given incoming power for monochromatic beams is attained when the beam is a perfect converging dipolar beam  \cite{bassett1986limit}. However, for propagating beams where the incident angles are limited to a hemisphere (i.e. $2\pi$ illumination - $0 \le \varphi \le 2 \pi$ and $\pi/2 < \theta \le \pi$ for a beam propagating along the $z$ axis) the properties of the dipolar beam in the focal region can be very different, although the total energy density at the focus can still be more than half of the maximum possible \cite{sheppard1994optimal}. 
The  vector properties of the light strongly affect the field polarization and intensity distribution near the focus of tightly focused vector beams\cite{dorn2003sharper,bauer2014nanointerferometric}
Our main goal here is to explore the field  properties of propagating spherical multipole beams within a framework based on helicity, angular momentum and symmetry \cite{fernandez2012helicity}.

\begin{figure}[h]
	\subfloat[Sectoral dipolar beam with well-defined eigenvalues $l=m= \sigma = 1$. \label{fig1a}]
		{\includegraphics[width=0.45\linewidth]{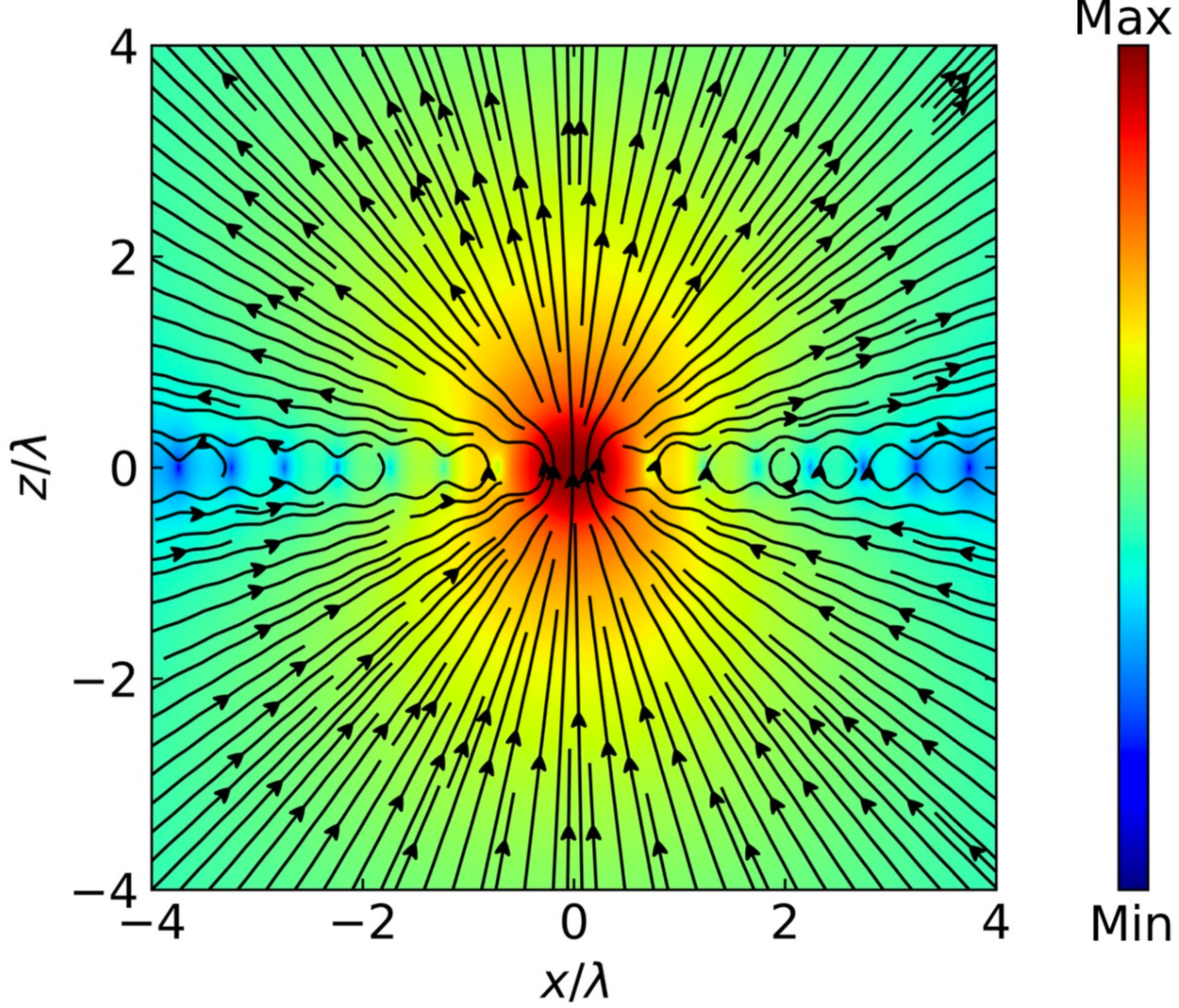}
		}\hfill
	\subfloat[Non-sectoral quadrupole beam with well-defined eigenvalues $l=2$ and  $m = \sigma = 1$.\label{fig1b}]{
		\includegraphics[width=0.45\linewidth]{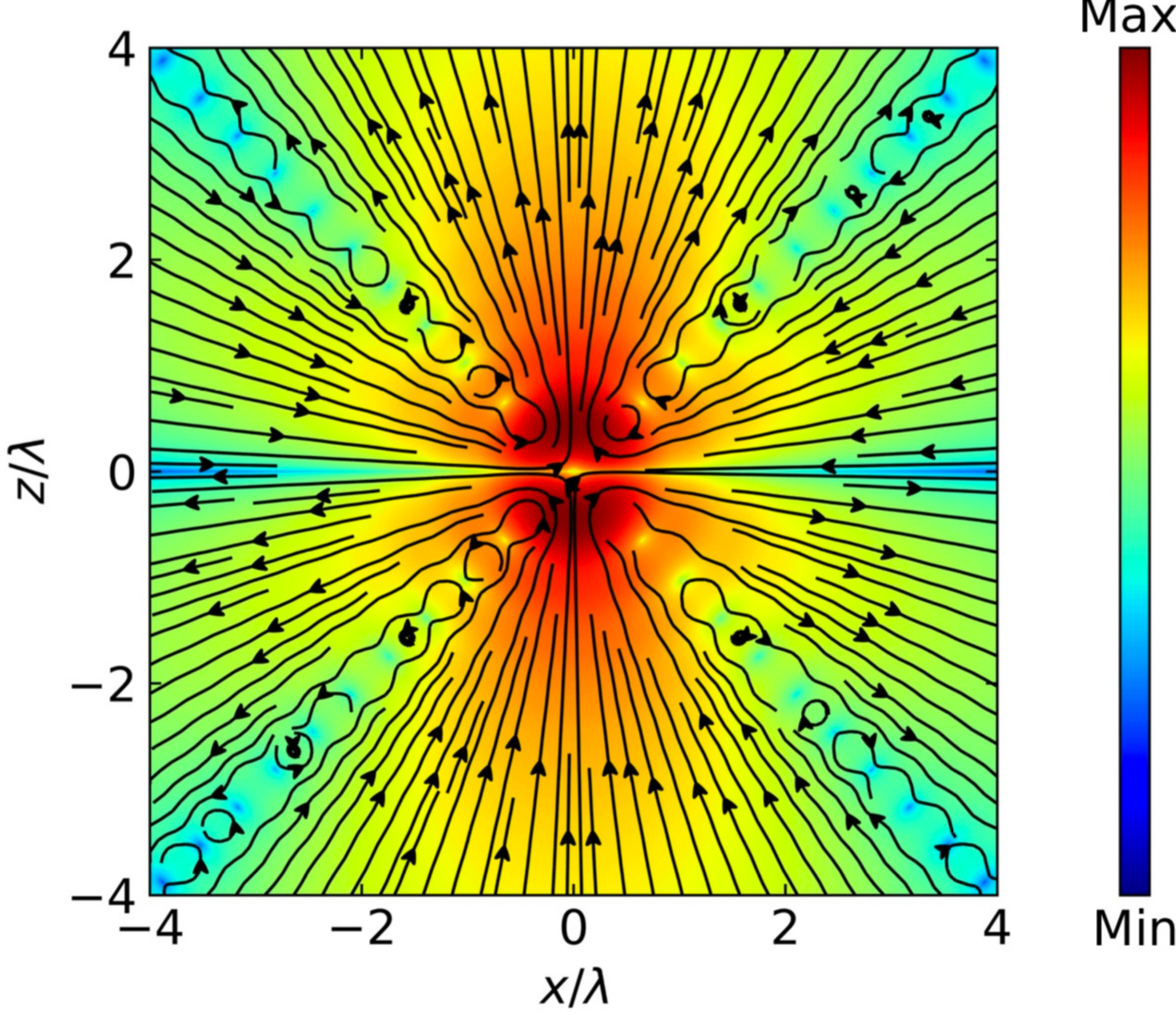}
		}
\caption{Poynting vector  streamlines projected on the $x-z$ plane, perpendicular to the focal plane $xy$. The colormap illustrates the modulus of the Poynting vector, $|{\bf{S}}|$, in the logarithm scale.}
\label{fig1}
\end{figure}

To this end,
we consider pure multipole beams (PMB) with well-defined handedness or helicity, $\sigma = \pm 1$ (we associate left polarized light with  $\sigma = +1$ positive  helicity  -handness-),  being simultaneous eigenvectors of the  squared total angular momentum,   $\bf{J}^2$, and its $z$-component,  $J_z$, with eigenvalues $l=1,2,\dots$ and  $m=-l, -l+1, \dots l-1, l$, respectively. 
For a beam propagating along the $z$-axis, the radial component of the Poynting vector far from the focus is assumed to be always negative for incoming light ($\pi/2 \le \theta \le \pi$) and always positive for outgoing ($0 \le \theta < \pi/2$). As we will see, this  condition of hemispherical illumination restricts the possible propagating multipole beams to the so-called ``sectoral'' multipoles with $m=\sigma l$. Figure \ref{fig1} illustrates two examples of sectoral and  non-sectoral beams.  Interestingly,
for $l>1$,   sectoral PMBs are vortex beams whose field vanishes on the propagation axis with a vortex topological  charge of $\sigma (l-1)$. In contrast,
 dipole beams  with well defined helicity ($l=1$, $m=\sigma$) concentrate the field at the focus with an energy density that is $2/3$ times the Bassett upper bound of passive energy concentration \cite{bassett1986limit}.   Dipole beams  are equivalent to mixed-dipole waves \cite{sheppard1994optimal} and can be seen as the sum of outgoing and incoming waves radiated from an electric and a magnetic dipole  located at the focus with spinning axes on the focal plane. We will see that,  although the helicity and the (dimensionless) $z$-component of the total angular momentum per photon ($m=\sigma l$) are fixed and well-defined in a PMB, the   ($z$-component)  SAM, $s_z$, and OAM, $l_z = \sigma l -s_z$, densities present a non-trivial spatial distribution.

\section{Beam expansion in vector spherical wavefunctions  in the helicity representation
}
Let us assume that a monochromatic light beam (with an implicit time-varying harmonic component $e^{-\im \omega t}$) propagates through an homogeneous  medium with real refractive index $n_h$ with a wave number $k=n_h \omega / c = 2 \pi n_h / \lambda_0$ (being $\lambda_0$ the light wavelength in vacuum).
To this end we will consider the expansion of the electric field, $\E$, of the incident focused beam in vector spherical wavefunctions (VSWFs), $\boldsymbol{\Psi}_{lm}^{\sigma}$,  with well defined helicity,  $\sigma = \pm 1$ \cite{olmos2019enhanced}:
\be
\E&=& \sum_{\sigma= \pm 1} \E_\sigma =  \sum_{\sigma= \pm 1} \left\{ \sum_{l=0}^{\infty} \sum_{m=-l}^{+l}  C_{lm}^{ \sigma} \boldsymbol{\Psi}_{lm}^{\sigma} \right\} 
\label{multipolar} 
\ee
where $\boldsymbol{\Psi}_{lm}^{\sigma}$ is defined as
\be
\boldsymbol{\Psi}_{lm}^{\sigma} &=& \frac{1}{\sqrt{2}} \left[ {\boldsymbol{N}}_{lm} +  \sigma {\boldsymbol{M}}_{lm}  \right] \\
{\boldsymbol{M}}_{lm} &\equiv & j_l(kr)\boldsymbol{X}_{lm}  \quad, \quad
{\boldsymbol{N}}_{lm} \equiv \frac{1}{k} \boldsymbol{\nabla} \times {\boldsymbol{M}}_{lm} \\
\boldsymbol{X}_{lm} &\equiv& \frac{1}{\sqrt{l(l+1)}} {\bf{L}} Y_l^m (\theta,\varphi)
\ee
Here, $\boldsymbol{M}_{lm}$ and $\boldsymbol{N}_{lm}$  are Hansen's multipoles \cite{borghese2007scattering},   $\boldsymbol{X}_{lm}$ denotes the vector spherical harmonic  \cite{jackson1999electrodynamics} , $ j_l(kr)$ are the spherical (well-defined at $r = 0$) Bessel functions,  $Y_l^m$ are the spherical harmonics and $ {\bf{L}} \equiv  \left\{ -\im \r \times \boldsymbol{\nabla}\right\} $ is the OAM operator.   The expansion coefficients $C_{lm}^{ \sigma}$ in the helicity basis are equivalent  to the so-called beam shape coefficients (BSCs) \cite{gouesbet1988light,neves2006exact}.  An explicit expression of the VSWFs in spherical polar coordinates (with unitary vectors $\hat{\boldsymbol{e}}_{r, \theta, \varphi}$) is given by
 \be
\boldsymbol{\Psi}_{lm}^{\sigma} &=& 
 {\hat{\boldsymbol{e}}_r} \ \im \sqrt{\frac{l(l+1)}{2}}\frac{j_l(kr)}{kr} Y_l^m \nonumber  +
  {\hat{\boldsymbol{e}}_\theta} \frac{1}{\sqrt{2l(l+1)}} \left\{ - 
  j_l(kr) \Big( \frac{ \sigma m}{\sin \theta} Y_l^m \Big)
 +  \im \tilde{j}_l(kr) \left(  \frac{\partial Y_l^m}{\partial \theta} 
\right)
\right\} \nonumber \\ 
&+&    {\hat{\boldsymbol{e}}_\varphi} \frac{1 }{\sqrt{2l(l+1)}}\left\{
- \tilde{j}_l(kr) \left(    \frac{  m}{\sin \theta} Y_l^m  
\right)
- \im  \sigma j_l(kr)  \left(  \frac{\partial Y_l^m}{\partial \theta} 
\right) \right\},
\ee
with
\be
\tilde{j}_l(kr)  &\equiv& j_{l-1}(kr)- l \frac{j_l(kr)}{kr} = (l+1) \frac{j_l(kr)}{kr} - j_{l+1}(kr).
\ee

 Let us recall that the multipoles $ \boldsymbol{\Psi}_{lm}^{\sigma} $ can be built following the standard rules of angular momentum addition \cite{borghese2007scattering,edmonds2016angular}
 as simultaneous eigenvectors of the square of the total angular momentum,   $\bf{J}^2$, and its $z$-component,  $J_z$, with 
 $\bf{J} = {\bf{L}} + \tensorS$  given by the sum of the OAM, $ {\bf{L}}$, and SAM, $\tensorS$, operators:
\be
\tensorS &\equiv& \im {\bf{I}} \times \ \quad, \quad  {\bf{L}} \equiv  \left\{ -\im \r \times \boldsymbol{\nabla}\right\} \label{SpinTensor} \\
\tensorS_i &=&  \hat{\boldsymbol{e}}_i \cdot \tensorS = \im \hat{\boldsymbol{e}}_i \times, 
\ee
where $\hat{\boldsymbol{e}}_{i=x,y,z}$ indicate unitary Cartesian vectors and 
 ${\bf{I}} $ is  the unit dyadic.
  
The multipoles $ \boldsymbol{\Psi}_{lm}^{\sigma} $ are then simultaneous eigenvectors of 
  $\bf{J}^2$ and  $J_z$  \cite{edmonds2016angular} as well as of the helicity operator \cite{fernandez2012helicity} $\boldsymbol{\Lambda} = (1/k) \boldsymbol{\nabla} \times$, i.e. :
 \be
{\bf{J}}^2 \boldsymbol{\Psi}_{lm}^{\sigma} &=& l(l+1) \boldsymbol{\Psi}_{lm}^{\sigma} \\
J_z \boldsymbol{\Psi}_{lm}^{\sigma} &=& m \boldsymbol{\Psi}_{lm}^{\sigma} \\
\boldsymbol{\Lambda} \boldsymbol{\Psi}_{lm}^{\sigma} &\equiv& \frac{1}{k} \boldsymbol{\nabla} \times  \boldsymbol{\Psi}_{lm}^{\sigma}  = \sigma \boldsymbol{\Psi}_{lm}^{\sigma} 
 \ee

\section{Poynting vector for propagating pure multipole beams} 

 In the helicity representation, the Poynting vector, $\boldsymbol{P}$, for a monochromatic optical field, when calculated using either the electric or the magnetic field, separates into right-handed and left-handed contributions, with no cross-helicity contributions \cite{aiello2015note}:
\be 
\boldsymbol{P} &\equiv& \frac{1}{2} \text{Re}\left\{ \E^* \times \H \right\} =  \sum_{\sigma= \pm 1}   \frac{\sigma}{2Z} \text{Im}\left\{ \E_\sigma^*  \times \E_\sigma \right\}   
=-\frac{1}{2Z} \text{Re} \Big(  \E_+^* \cdot \tensorS \E_+ -  \E_-^* \cdot \tensorS \E_-\Big). \label{PandS}
\ee
where $\tensorS$ is the spin tensor defined in Eq. \eqref{SpinTensor} and
\be
\H &=&   -\frac{\im}{Z k} \boldsymbol{\nabla} \times \Big( \E_+ + \E_-\Big)  =  -\frac{\im}{Z} \Big( \E_+ - \E_-\Big)\ee
with $Z= 1/ (\epsilon_0 n_h c) = \sqrt{\mu_0/\epsilon_0\epsilon_h}$. 
This is an interesting result showing that 
for beams with well defined helicity $\sigma$,  the $z$-component of the spin is simply proportional to the proyection of the Poynting vector on the propagation axis. 

In the far-field, where
\be
 \lim_{ kr \to \infty}  j_j \sim \frac{\sin \big(kr-l \pi/2\big)}{kr}
,
\ee
the Poynting vector
for a pure multipole beam (PMB)  (with $l,m,\sigma$) is given by 
 \be
  \boldsymbol{P}_{lm}^\sigma \sim  \frac{ \sigma m \left| C_{lm}^\sigma\right|^2 
}{{\bf{4}}Z( kr) ^2 \ l(l+1)} \Bigg[  
  \frac{ 1 }{ \sin \theta}  \frac{\partial \left|Y_l^m\right|^2}{\partial \theta} 
 \Bigg] {\hat{\boldsymbol{e}}_r}. \label{PPMBff1}
\ee
Integrating over all incoming angles ($\pi/2 \le \theta \le \pi$), the total incoming power, $P_W$, for a pure multipole beam is then given by
\be
P_W(l,m,\sigma)&=&- r^2\int_0^{2\pi} d \varphi\int_{\pi/2}^\pi \sin\theta d \theta   \ \boldsymbol{P}_{lm}^\sigma \cdot  {\hat{\boldsymbol{e}}_r} \nonumber   =  2 \pi \frac{ \sigma m \left| C_{lm}^\sigma\right|^2 
}{4 Z k^2 \ l(l+1)}  
\left|Y_l^m\right|^2_{\theta=\pi/2} 
\ee
which, from the parity relation of the spherical harmonics and associated Legendre functions, is zero for  $l+m$ odd. 
This implies  that the total amount of power that is carried by   beams with $l+m$ odd  is identical to zero, even though they may present incoming and outgoing  Poynting vectors (notice that $P_W$ is the actual total power flowing through the focal plane). We shall consider incoming beams whose far-field Poynting vector points towards (outwards)  the focus 
for all incoming angles $\pi/2 < \theta \le \pi$  ( outgoing angles $0 \le \theta < \pi/2$ ), i.e. whose far-field Poynting vector only changes sign  at $\theta=\pi/2$:
\begin{align} \label{SECTORAL}
{\hat{\boldsymbol{e}}_r} \cdot \boldsymbol{P}_{lm}^\sigma <0,   \quad \text{for} \quad \pi/2 \le \theta \le \pi. && \text{and} &&  {\hat{\boldsymbol{e}}_r} \cdot \boldsymbol{P}_{lm}^\sigma >0,   \quad \text{for} \quad 0 \le \theta \le \pi/2 .
\end{align} 
%
\begin{figure}[h]
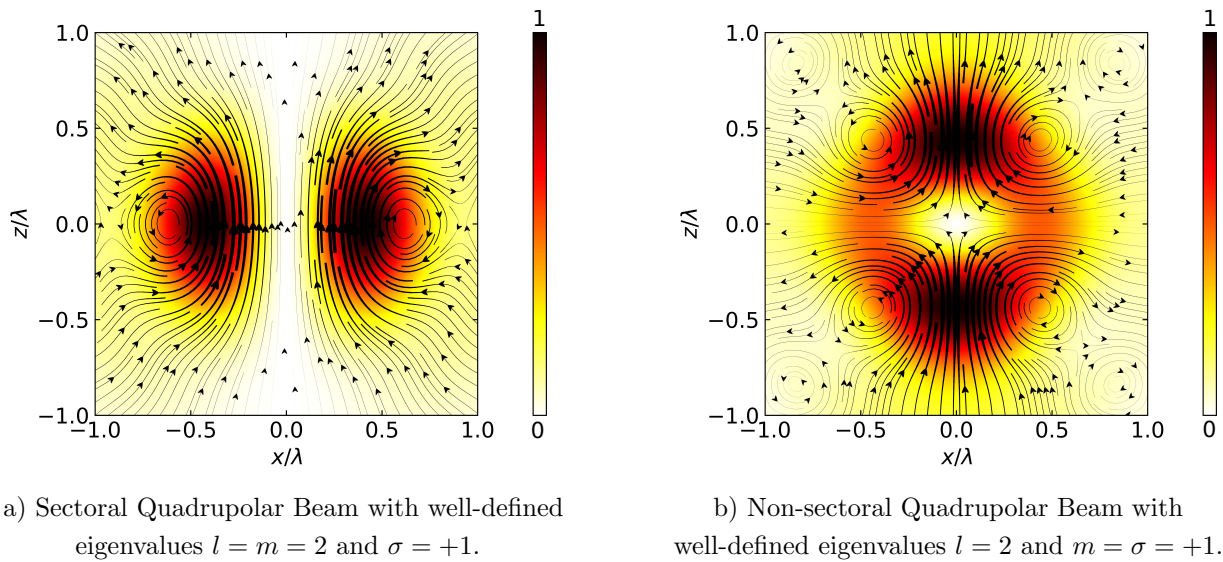

	\subfloat[Sectoral Quadrupolar Beam with well-defined eigenvalues $l=m=2$ and $\sigma = +1$. \label{fig2a}]
		{\includegraphics[width=0.45\linewidth]{FIGURE_2A}
		}\hfill
	\subfloat[Non-sectoral Quadrupolar Beam with well-defined eigenvalues $l=2$ and $m=\sigma = +1$.\label{fig2b}]{
		\includegraphics[width=0.45\linewidth]{FIGURE_2B}
		}
\caption{Poynting vector  streamlines projected on the $x-z$ plane, perpendicular to the focal plane $xy$. The intensity color map corresponds to the field intensity normalised to the maximum field intensity for each case $ I = |\E|^2 / |\E_{\rm{max}}|^2 $. The maximum value is in both cases approximately given by $|\E_{\rm{max}}|^2 = 0.4 \left|\E_{1,\sigma}^{\sigma} (\r=0) \right|^2  $, being $\left|\E_{1,\sigma}^{\sigma} (\r=0) \right|^2$ the intensity at the focal plane in the (sectoral) dipolar case. }
\label{fig2}
\end{figure}

Since  $ {\hat{\boldsymbol{e}}_r} \cdot \boldsymbol{P}_{lm}^\sigma$ changes sign at the maxima and minima of  $\left| Y_l^m(\theta, \phi)\right|^2$,  which has $l-|m|$ distinct zeros in the interval $(0 < \theta < \pi)$, the only possible PMBs are the so-called `sectoral' multipoles with $m=\sigma l$. As an illustrative example, Fig. \ref{fig2} shows the Poynting vector map of   sectoral and a non-sectoral quadrupole beams with $l=2$.  From these plots it is straightforward to notice that the far-field Poynting streamlines of the sectoral beam, corresponding to Fig. \ref{fig2}a), satisfies Eq.\eqref{SECTORAL},  while the non-sectoral, illustrated in Fig. \ref{fig2}b), does not fulfil the propagating PMB criterium.

\section{Sectoral multipole beams with well defined helicity}
Taking into account that 
\be
Y_l^{m=\sigma l} = \frac{(-\sigma)^l}{2^l l! } \sqrt{\frac{(2l+1)!}{4\pi}} \sin^l\theta e^{\im \sigma l \varphi},
\ee
it can be shown  that the electric field of a sectoral PMB (with total incoming power $P_W$) in spherical polar coordinates [$E_\beta$ with $\beta=r, \theta, \varphi$] can be written as
{
\be
\E_{l,\sigma l}^{ \sigma}= C_{l,\sigma l}^{ \sigma} \boldsymbol{\Psi}_{l, \sigma l}^{\sigma} &=&
 E_0   (-\sigma)^l \sin^{l-1}\theta e^{\im \sigma l \varphi}  \Big[
  {\hat{\boldsymbol{e}}_r} \ \im (l+1) \frac{j_l}{kr}  \sin\theta \nonumber +
  {\hat{\boldsymbol{e}}_\theta} \left\{ - 
  j_l 
 +  \im \tilde{j}_l\cos\theta \right\}  \\ &+&
    {\hat{\boldsymbol{e}}_\varphi} \left\{
-  \sigma \tilde{j}_l- \im  \sigma j_l  \cos\theta  \right\}
\Big] \label{ESPMB} \ee with
\be E_0 &\equiv& k \sqrt{\frac{ l \ Z P_W}{ \pi}}. \ee
In the spherical basis 
 \be
 \xim_{+1} = -\frac{\hat{\boldsymbol{e}}_x+\im  \hat{\boldsymbol{e}}_y}{\sqrt{2}}, \quad
 \xim_0=  \hat{\boldsymbol{e}}_z, \quad
  \xim_{-1}= \frac{\hat{\boldsymbol{e}}_x-\im  \hat{\boldsymbol{e}}_y}{\sqrt{2}},
  \ee
this reads as
  \be
&\E_{l,\sigma l}^{ \sigma}&=
 E_0  (-\sigma)^l  \sin^{l-1}\theta  e^{\im \sigma l \varphi}  \frac{1}{\sqrt{2}}\Bigg[
\xim_{+1}  e^{-\im \varphi} \Big(-
   \im j_{l+1} \sin^2\theta +(\sigma +1)\left[j_l  \cos\theta - \im  \tilde{j}_l \right]
 \Big)\nonumber \\ &+&  \xim_0 
 \sqrt{2}  \Big(  \im j_{l+1} \cos\theta
+
  j_l 
   \Big) \sin\theta + \xim_{-1} e^{\im \varphi}
    \Big( \im j_{l+1}\sin^2\theta
    +(\sigma -1)\left[j_l  \cos\theta - \im  \tilde{j}_l \right] \Big)
\Bigg]. \ee
}
The behaviour near the $z$-axis, i.e. for $\theta \lesssim \pi$ and $\theta \gtrsim 0$,
\be
\E_{l,\sigma l}^{ \sigma} &\sim &  E_0  (-\sigma)^l  \sin^{l-1}\theta
\frac{2\sigma}{\sqrt{2}} e^{\im \sigma (l-1) \varphi} \Big[
 \frac{z}{|z|} j_l - \im \tilde{j}_l\Big] \xim_{\sigma} \\   \lim_{ kr \to \infty} \E_{l,\sigma l}^{ \sigma} & \sim &
-  E_0  (-\sigma)^l  \sin^{l-1}\theta \frac{2 \im \sigma}{\sqrt{2}}  e^{\im \sigma (l-1) \varphi} \  \frac{e^{\im kz}}{k |z|}  \xim_{\sigma} \ee
shows that,
for $l > 1$, the sectoral PMBs are vortex beams whose field vanishes on the propagation axis with a vortex topological charge of $\sigma(l-1)$. Figure \ref{fig2a} shows the Poynting vector lines of a sectoral quadrupolar beam ($l=2,m=2,\sigma=+1$) which present a "doughnut-like" field intensity pattern around the focus.  

\begin{figure}[h]
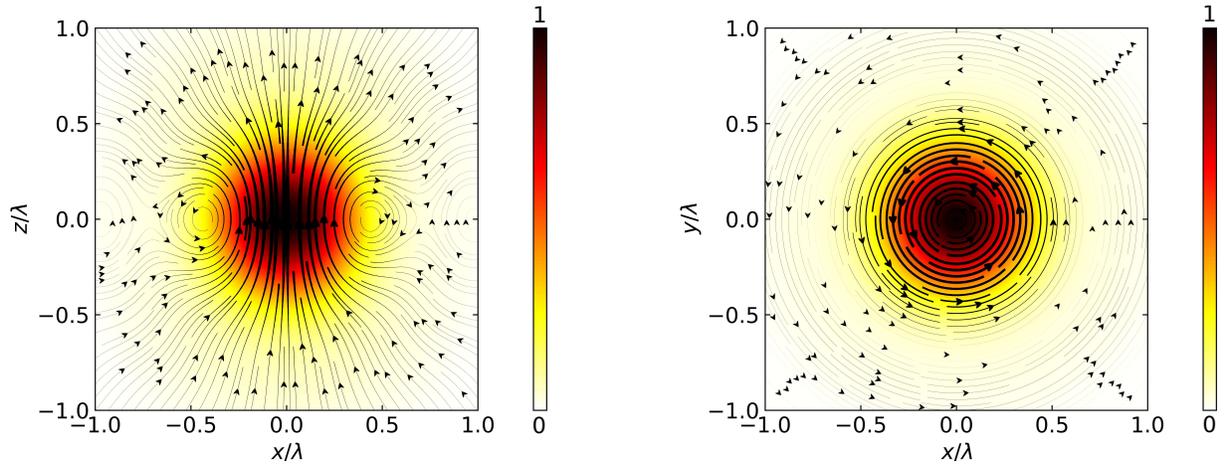

	\subfloat[$\boldsymbol{P}_{11}^{+1}$ projected on the $x-z$ plane,  with  $y = 0$.  \label{fig3a}]
		{\includegraphics[width=0.45\linewidth]{FIGURE_3A}
		}\hfill
	\subfloat[$\boldsymbol{P}_{11}^{+1}$ projected on the focal plane, with $z = 0$.  \label{fig3b}]{
		\includegraphics[width=0.45\linewidth]{FIGURE_3B}
		}
\caption{Projected flow lines of the Poynting vector, $ \boldsymbol{P}_{1,1}^{+1}$ and field intensities for a dipole beam.  The intensity color map corresponds to the  field intensity normalised to the maximum field intensity at the center $ I = |\E|^2 /|\E_{\rm{max}}|^2 $, where $|\E_{\rm{max}}|^2 = 8|\E_0|^2 / 9 =  |C^{+1}_{11}|^2 / 12 \pi $.}
\label{fig3}
\end{figure}

For pure dipolar beams with $l=1$,  the field given by  Eq. \eqref{ESPMB} is identical to the first term of the expansion of a circularly polarized plane wave around the origin \cite{jones2015optical,Zaza2019Size}.  The field at the focus, $\r=0$,  given by
\be
\E_{1,\sigma}^{\sigma} (\r=0) &=& \im \frac{2 \sqrt{2}}{3}  E_0  \ \xim_\sigma,
\ee
 is circularly polarized on the focal plane.
 The ratio between the electric energy density at the focal point and the total incoming power  is then given by
\be
\frac{U}{P_W} &=& \frac{\epsilon_0 \epsilon}{4 P_W} \left|\E_{1,\sigma}^{\sigma} (\r=0) \right|^2 =
\frac{2}{3} \left\{\frac{k^2}{3 \pi} \frac{n_h}{c}\right\} \ee
i.e. $2/3$ of the Bassett upper bound. The field intensities and Poynting vector lines for a dipole beam are shown in Fig. \ref{fig3}. As it can be seen the intensity is concentrated in a volume $\sim (\lambda/2)^3$, as expected for a diffraction limited beam with an interesting  toriodal flow of the  Poynting vector around the focus. It is worth mentioning that steady-state toroidal current flow of the Poynting vector of highly focused beams  was already demonstrated experimentally and explained with an approximate field-model \cite{roichman2008influence}.


\subsection{Dipolar beams and time-reversal of a Huygens's source}

Dipole beams are equivalent to mixed-dipole waves \cite{sheppard1994optimal} and can be seen as the sum of outgoing and incoming waves radiated from an electric dipole,  $\p^{(\sigma)}$, and an equivalent magnetic dipole, $\m^{(\sigma)}$, located at the focus. These electric and magnetic point sources are known as Huygens' sources and their direct, outgoing, emission present a peculiar asymmetric distribution of the radiation pattern \cite{sheppard1994optimal,gomez2011electric,geffrin2012magnetic}.  As we show in the Appendix \ref{Green}, the field of a dipole beam  ( Eq. \eqref{ESPMB} with $l=1$) can be rewritten as
\be
 \E_{1,\sigma}^\sigma &=&  - 2 \im \frac{k^2}{\epsilon_0 \epsilon} 
\text{Im}\left\{G_{ee}(\r)\right\} {\p^{(-\sigma)}}^* 
+ 2 k^2 Z \text{Im}\left\{G_{em}(\r)\right\}  {\m^{(-\sigma)}}^*, \label{TRC}
\ee
where $G_{ee}(\r)$ is the outgoing dyadic Green function, $G_{em} \equiv (1/k)  \boldsymbol{\nabla} \times G_{ee}$ and
\begin{align}
\frac{\p^{(\sigma)}}{\epsilon_0 \epsilon_h} = \frac{ {\p^{(-\sigma)}}^* }{\epsilon_0 \epsilon_h}  = - \frac{3\pi}{k^3} \E_{1,\sigma}^\sigma(\r=0), && \text{with} &&
 \m^{(\sigma)} = {\m^{(-\sigma)}}^* =- \frac{3\pi}{k^3} \H_{1,\sigma}^\sigma(\r=0).
\end{align}
Interestingly, for an incoming beam with helicity $\sigma$, Eq. \eqref{TRC}  is exactly the time-reversed electric field radiated by the Huygens' source (with helicity
$-\sigma$) as it would be obtained in a time-reversal mirror cavity \cite{carminati2007theory}.

\section{Linearly polarized dipolar beams}
\begin{figure*}[h]
\includegraphics[width=1 \textwidth]{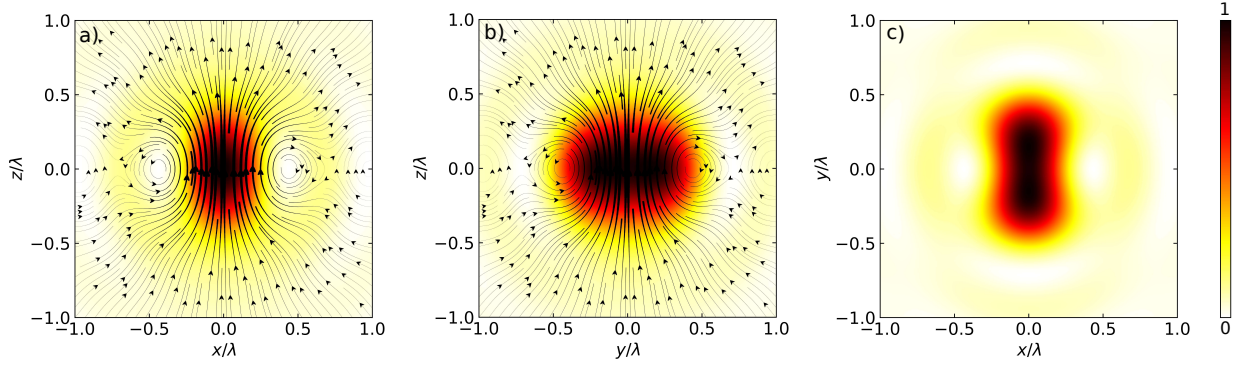}
\caption{Projected Poynting vector and colormap of the intensity for linarly polarized dipolar beams (a) on the $x-z$ plane, (b) on the $y-z$ plane and (c) on the $x-y$ focal plane. The streamlines of the Poynting vector are  normal to the focal plane.  }\label{fig4}
\end{figure*}

Focused linearly polarized dipolar beams  can be built by combining two well-defined helicity beams with opposite signs from Eq.\eqref{ESPMB}, i.e. 
\begin{align}
\E_x(\r) = \frac{1}{\sqrt{2}} \left(\E_{1,-1}^{-1}(\r) - \E_{1,1}^1(\r) \right) && \text{and} &&  \E_y(\r) = \frac{\im}{\sqrt{2}}\left(\E_{1,-1}^{-1}(\r) + \E_{1,1}^1(\r) \right).\end{align}
The field at the focus is linearly polarized on the focal plane
\be
\E_{x,y}(\r=0) &=& \im \frac{2 \sqrt{2}}{3}  E_0 \ \hat{\boldsymbol{e}}_{x,y} .
\ee
The Poynting vector, obtained  by inserting Eq.\eqref{PPMB} into Eq.\eqref{PandS} including both helicities in the summation,  present the characteristic toroidal flow around the focus but, as expected, the intensity pattern at the focal plane is not longer axially symmetric (see Fig. \ref{fig4}).

\section{Spin and Orbital angular momenta of pure multipolar beams}

Let us now consider  the separation of the total angular momentum of the  sectoral PMB  field into SAM and OAM. 
Since the PMBs are eigenvectors of the $z$-component of the total angular momentum, we may define its  (constant) density as
\be
\frac{ \left\{\E_{l,\sigma l}^\sigma \right\}^*  \cdot J_z \E_{l,\sigma l}^\sigma}{\left|{\E_{l,\sigma l}^\sigma}\right|^2
} = 
m = \sigma l = \frac{ \left\{\E_{l,\sigma l}^\sigma \right\}^* \cdot \Big(L_z + S_z \Big) \E_{l,\sigma l}^\sigma}{\left|{\E_{l,\sigma l}^\sigma}\right|^2} = l_z(\r) + s_z(\r)
\ee
where we have introduced $l_z$ and $s_z$ as the OAM and SAM densities, respectively (although $L_z$ and $S_z$ are not proper angular-momentum operators\cite{van1994spin}).  For fields with well defined helicity $\sigma$,  the $z$ component of the SAM can be calculated from the projection of the Poynting vector on the beam axis, i.e.  from Eqs. \eqref{PandS} and \eqref{PPMB},  or, alternatively,
taking into account that the vectors $\boldsymbol{\xi}_\mu$ are eigenfunctions of $S_z$ with eigenvalue $\mu$:
\be
\left\{\E_{l,\sigma l}^\sigma \right\}^*   \cdot S_z \E_{l,\sigma l}^\sigma &=& 2 Z \sigma \boldsymbol{P}_{l,\sigma l}^\sigma \cdot {\hat{\boldsymbol{e}}_z} = \sum_{\mu=+1,0,-1} 
\mu\left| { \E_{l,\sigma l}^\sigma \cdot \xim_{\mu} }\right|^2 \nonumber \\ &=&
2 \sigma  \frac{|E_0|^2}{Z}  \sin^{2l-2}\theta \left\{
  \left[\left|\tilde{j}_l\right|^2    +     \left|j_l\right|^2  \right] \cos^2\theta
  + \frac{l+1}{kr}\left\{  j_l \tilde{j}_l  \right\} \sin^2\theta
\right\} 
\ee

\begin{figure}[t]
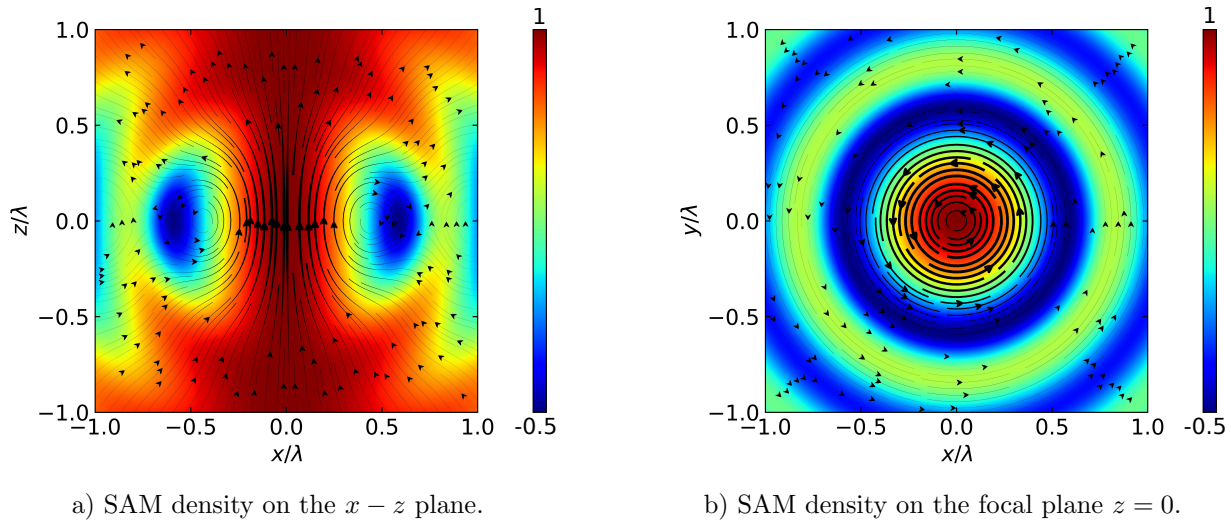

	\subfloat[SAM density  on the $x-z$ plane.  \label{fig5a}]
		{\includegraphics[width=0.45\linewidth]{FIGURE_5A}
		}\hfill
	\subfloat[SAM density on the focal plane $z=0$.   \label{fig5b}]{
		\includegraphics[width=0.45\linewidth]{FIGURE_5B}
		}
\caption{Projected Poynting vector, $ \boldsymbol{P}_{1,1}^{+1}$, flow lines and  color map of the $z$-component of the SAM density for a dipole beam. The toroidal structure can be easily inferred from the sign of the spin density in both planes.}
\label{fig5}
\end{figure}

Figure \ref{fig5} illustrates the non-trivial behaviour of the SAM for a dipole beam ($\sigma=+1$)  on two perpendicular planes in the focal volume.
As shown in Fig \ref{fig5a}, the spin changes sign inside the torus defined by the Poynting vector lines (which is simply due to the change  of sign of the $z$-component of the Poynting vector in the torus). In these (blue) regions, the $z$-component of the OAM $l_z$ is larger than the $z$-component  of the total momentum $l=1$ (since $l_z =  l - s_z$). This is often referred to in the literature as the super-momentum regime \cite{bekshaev2015transverse,araneda2019wavelength,olmos2019asymmetry}.


\section{Concluding remarks}

Pure multipolar  beams are exact analytical solutions of Maxwells equations with well defined angular momentum properties that can be extremely useful to understand and model different phenomena associated to light-matter interactions under strongly focused beams without requiring sophisticated and cumbersome numerical calculations. We have shown that dipole beams could be generated by time-reversal techniques \cite{carminati2007theory} as the time reversal of a Huygens' source. 
The steady-state toroidal Poynting vector  flow in focused beams, which was already demonstrated experimentally and explained with  approximated beam models \cite{roichman2008influence}, 
appears here in a natural way in the exact analytical solution for the dipole beam. The peculiar distribution of the SAM and OAM around the focus could lead to some interesting angular momentum transfer phenomena to small asymmetric and/or absorbing particles trapped by circularly polarized highly focused beams \cite{ruffner2012optical}. 
Higher sectoral multipole vortex beams, with well-defined total angular momentum and its projection on the propagation axis could also be useful to understand scattering and AM phenomena in vortex beams  \cite{chen2013dynamics}.  The peculiar properties of the fields in the focus of highly focused beams may also be relevant to understand the  light-induced emergence of cooperative phenomena in colloidal suspensions of nanoparticles \cite{kudo2018single,delgado2018emergence}.

\section*{Funding}
This research was supported by the Spanish Ministerio
de Econom\'ia y Competitividad (MICINN) and European
Regional Development Fund (ERDF) [FIS2015-69295-C3-3-P] and by the Basque Dep. de Educaci\'on  [PI-2016-1-0041] and PhD Fellowship (PRE- 2018-2-0252).

\appendix

\section{Poynting vector for sectoral multipole beams}
Using Eq. \eqref{PandS} together with\eqref{ESPMB}, the Poynting vector for a sectoral multipole beam is given by:
\be
 \boldsymbol{P}_{l,\sigma l}^\sigma= 
 k^2 P_W \frac{ l   }{ \pi} \sin^{2l-2}\theta
   \Bigg[ {\hat{\boldsymbol{e}}_r} 
  \left[\left|\tilde{j}_l\right|^2    +     \left|j_l\right|^2  \right] \cos\theta
&+&{\hat{\boldsymbol{e}}_\theta} \left\{-  \frac{ l+1  }{kr}
\left\{  j_l \tilde{j}_l  \right\} \sin\theta
\right\} 
 \nonumber \\
&+&  {\hat{\boldsymbol{e}}_\varphi}
 \left\{    
 \frac{(l+1)}{kr} \left[ \left| j_l\right|^2  \sigma \sin\theta
 \right]
\right\} 
 \Bigg] \label{PPMB}
\ee

\section{Field of a sectoral dipole beam in terms of the Green tensors}
\label{Green}
The electric field radiated by an electric dipole, $\p^{(\sigma)}$, and a magnetic dipole, $\m^{(\sigma)}$,  located at the origin of coordinates can be written as
\be
\E_d &=& k^2 \G_{ee}(\r) \frac{\p^{(\sigma)}}{\epsilon_0\epsilon_h} + k^2 \G_{em} \im Z \m^{(\sigma)}
= \frac{k^2}{\epsilon_0\epsilon_h}  \left(  \G_{ee}(\r) \p^{(\sigma)} + \im \frac{n_h}{c} \G_{em}  \m^{(\sigma)} \right) \ee
with \be
\G_{ee}(\r) \p^{(\sigma)}&=& -\im \frac{k}{4 \pi } \left\{ h_0(kr) \frac{(\r \times \p^{(\sigma)})\times \r}{r^2} + \frac{h_1(kr)}{kr} \left(\frac{3(\r \cdot \p^{(\sigma)})\r}{r^2} -\p^{(\sigma)}\right) \right\},  \\
\G_{em}(\r) \m^{(\sigma)}&=& -\im \frac{k}{4 \pi } h_1(kr) \frac{\r \times \m^{(\sigma)}}{r}.
\ee 
In spherical polar coordinates,
\be
\E_d = \begin{pmatrix} E_r \\ E_\theta \\ E_\varphi \end{pmatrix} &=& -\im \frac{k^3}{4 \pi \epsilon_0\epsilon_h}
\begin{pmatrix}
2\frac{h_1(kr)}{kr}  &0 & 0 \\
0 & \tilde{h_1}(kr) & 0 \\
0 & 0 & \tilde{h_1}(kr) 
\end{pmatrix}
\begin{pmatrix} p_r \\ p_\theta \\ p_\varphi \end{pmatrix} \nonumber \\
&& 
+ \frac{k^3}{4 \pi} Z  h_1(kr) 
\begin{pmatrix}
0 &0 & 0 \\
0 & 0 & -1 \\
0 & 1& 0
\end{pmatrix}
\begin{pmatrix} m_r \\ m_\theta \\ m_\varphi \end{pmatrix} 
\ee
where $h_1$ are the outgoing Hankel functions, $j_1 =(h_1+h_1^*)/(2\im)$, and 
\be
\tilde{h}_1(kr)  &\equiv& h_{0}(kr)-  \frac{h_1(kr)}{kr}.
\ee

Assuming
\be
\frac{\p^{(\sigma)}}{\epsilon_0 \epsilon_h} &=& - \frac{3\pi}{k^3} \E_{1,\sigma}^\sigma(\r=0) =
 -\im  \frac{4\pi}{k^3}   \frac{ E_0}{\sqrt{2}} \ \xim_\sigma
=  \im \sigma \frac{4\pi}{k^3}  \frac{E_0}{2} e^{\im \sigma \varphi} \begin{pmatrix} \sin \theta \\ \cos\theta \\ \im \sigma \end{pmatrix} \\
 \m^{(\sigma)} &=& - \frac{3\pi}{k^3} \H_{1,\sigma}^\sigma(\r=0)  =   \frac{3\pi}{k^3}  \frac{\im \sigma}{Z}  \E_{1,\sigma}^\sigma(\r=0),  \ee
It is easy to show that
\be
\E_d = \E_{1,\sigma}^\sigma &=& - 2 \im \frac{k^2}{\epsilon_0 \epsilon} 
\text{Im}\left\{\G_{ee}(\r)\right\} \p^{(\sigma)} 
+ k^2 Z \text{Im}\left\{\G_{em}(\r)\right\}  \m^{(\sigma)} \nonumber \\
&=& - 2 \im \frac{k^2}{\epsilon_0 \epsilon} 
\text{Im}\left\{\G_{ee}(\r)\right\} {\p^{(-\sigma)}}^* 
+ k^2 Z \text{Im}\left\{\G_{em}(\r)\right\}  {\m^{(-\sigma)}}^* ,
\ee
which corresponds to Eq. \eqref{TRC} and the extension of Eq. (8) of Carminati {\em et al.} \cite{carminati2007theory} for electric and magnetic sources.

\bibliography{Focused_Beams}

\end{document}